\documentclass[nofootinbib,superscriptaddress,a4paper,twocolumn, floatfix]{revtex4-2}
\usepackage[english]{babel}
\usepackage[utf8]{inputenc}

\usepackage{amsthm}
\usepackage{mathtools}
\usepackage{physics}
\usepackage{xcolor}
\usepackage{graphicx}
\usepackage[left=23mm,right=13mm,top=35mm,columnsep=15pt]{geometry} 
\usepackage{algpseudocode}

\usepackage{adjustbox}
\usepackage{placeins}
\usepackage[T1]{fontenc}
\usepackage{lipsum}
\usepackage{csquotes}

\usepackage[pdftex, pdftitle={Article}, pdfauthor={Author}]{hyperref} 
\usepackage{xcolor}
\hypersetup{
    colorlinks,
    linkcolor={red!50!black},
    citecolor={blue!50!black},
    urlcolor={blue!80!black}
}
\usepackage{booktabs}
\usepackage{multirow}
\usepackage{listings}
\usepackage{xcolor}

\definecolor{codegreen}{rgb}{0,0.6,0}
\definecolor{codegray}{rgb}{0.5,0.5,0.5}
\definecolor{codepurple}{rgb}{0.58,0,0.82}
\definecolor{tqblue}{HTML}{08293d}
\definecolor{backcolour}{HTML}{fefdf5}

\lstdefinestyle{mystyle}{
    backgroundcolor=\color{backcolour},   
    commentstyle=\color{codegreen},
    keywordstyle=\color{magenta},
    numberstyle=\tiny\color{codegray},
    stringstyle=\color{codepurple},
    basicstyle=\ttfamily\footnotesize\color{tqblue},
    breakatwhitespace=false,         
    breaklines=true,
    postbreak=\mbox{\textcolor{magenta}{$\hookrightarrow$}\space},                 
    captionpos=b,                    
    keepspaces=true,                 
    numbers=left,                    
    numbersep=5pt,                  
    showspaces=false,                
    showstringspaces=false,
    showtabs=false,                  
    tabsize=2
}

\newcommand{\No}{\ensuremath{N_\text{o}}}

\newcommand{\Ne}{\ensuremath{N_\text{e}}}
\newcommand{\Nq}{\ensuremath{N_\text{q}}}
\newcommand{\N}[1]{\ensuremath{N_\text{#1}}}

\newcommand{\pairwfn}{{\ensuremath{\Psi_\text{SP}}}}
\newcommand{\lr}[1]{\ensuremath{\left( #1 \right)}}

\newcommand{\shiftleft}[2]{\makebox[0pt][r]{\makebox[#1][l]{#2}}}

\begin{document}

\title{Optimized Low-Depth Quantum Circuits for Molecular Electronic Structure using a Separable Pair Approximation}

\author{Jakob~S.~Kottmann}
\email[E-mail:]{jakob.kottmann@utoronto.ca}
\affiliation{{Chemical Physics Theory Group, Department of Chemistry, University of Toronto, Canada.}}
\affiliation{Department of Computer Science, University of Toronto, Canada.}

\author{Alán Aspuru-Guzik}
\email[E-mail:]{aspuru@utoronto.ca}
\affiliation{{Chemical Physics Theory Group, Department of Chemistry, University of Toronto, Canada.}}
\affiliation{Department of Computer Science, University of Toronto, Canada.}
\affiliation{Vector Institute for Artificial Intelligence, Toronto, Canada.}
\affiliation{Canadian  Institute  for  Advanced  Research  (CIFAR)  Lebovic  Fellow,  Toronto,  Canada}

\date{\today} 

\begin{abstract}
We present a classically tractable model that leads to optimized low-depth quantum circuits leveraging separable pair approximations.
The obtained circuits are well suited as a baseline circuit for emerging quantum hardware and can, in the long term, provide significantly improved initial states for quantum algorithms. The associated wavefunctions can be represented with linear memory requirement which allows classical optimization of the circuits and naturally defines a minimum benchmark for quantum algorithms.
In this work, we employ directly determined pair-natural orbitals within a basis-set-free approach. This leads to accurate representation of the one- and many-body parts for weakly correlated systems and we explicitly illustrate how the model can be integrated into other quantum algorithms for stronger correlated systems.
\end{abstract}

\maketitle

The electronic structure problem of quantum chemistry is one of the main anticipated applications for future quantum computers.~\cite{cao2019quantum, reiher2017elucidating}
The core problem is to extract, primarily low-lying, eigenenergies from electronic Hamiltonians as well as preparing the corresponding eigenstates. Algorithms, like the quantum phase estimation~\cite{aspuru2005simulated} or imaginary time evolution~\cite{motta2020determining, sun2021quantum} are promising candidates for application on future quantum computers. Since the first proposals, some of those algorithms have been improved significantly with regards to their resource requirements and estimated runtimes~\cite{babbush2018low-depth, vonburg2021quantum,lee2020efficient}, but their expected applicability remains out of reach for near and medium term devices.
Variational algorithms have been originally proposed~\cite{peruzzo2014variational, mcclean2016theory} as an applicable class of algorithms for those devices and as a potential bridging technology until scalable fault-tolerant hardware becomes available.
As the success probability (runtime) of phase estimation (imaginary time evolution) based algorithms depends on the overlap of the initial state with the targeted eigenstate, variational algorithms might also play a role in the long term by providing improved initial states.\\
Variational algorithms, usually aim to  prepare wavefunctions directly by a parametrized quantum circuit and measure the associated energies as expectation values. The parameters of the quantum circuit (\textit{e.g.} the angles of Eq.~\eqref{eq:ucc_gate_general}) are then optimized successively by a classical optimizer via the variational principle.
In its original form, which is most common in quantum chemical applications, the objective of the optimization is the plain expectation value of the electronic Hamiltonian. Detailed introductions to variational quantum algorithms can be found in recent reviews~\cite{bharti2021noisy, cerezo2020variational}, original articles~\cite{mcclean2016theory} or in recent works on unitary coupled-cluster with hands-on code examples~\cite{kottmann2020feasible, tequila}. The reviews~\cite{anand2021quantum, cao2019quantum, mcardle2020quantum} provide a general overview over quantum algorithms for quantum chemistry.
\\
The construction of suitable parametrized circuits for variational algorithms is a vibrant research topic, where \textit{physically inspired} models like unitary coupled-cluster where part of the initial proposals~\cite{mcclean2016theory} and remain a source of inspiration for current approaches.
Standard unitary coupled-cluster approaches such as UCCSD suffer from high gate and parameter counts that both scale with quartic cost in the number of orbitals. In addition, the corresponding quantum circuits show, once compiled into primitive one- and two-qubit gates,  high gate counts and depths even for small systems with a relatively small number of variational parameters. 
The standard approach defines the whole unitary operation through a single exponential generated by a sum over primitive electronic excitation operators (as defined in Eq.~\eqref{eq:ucc_generator_general}).
This unitary operation is then decomposed into primitive excitation unitaries (such as Eq.~\eqref{eq:ucc_gate_general}) employing for example the Trotter decomposition.
Most modern approaches abandoned this formulation over a single exponential and instead follow a factorized~\cite{evangelista2019exact} approach where the quantum circuit is constructed as a product of primitive excitations by various strategies such as adaptive circuit construction~\cite{grimsley2019adaptive, kottmann2020feasible, tang2021qubit} as well as Lie-algebraic~\cite{Izmaylov2020order} and empirical~\cite{grimsley2019trotterized} techniques.
Although significantly reduced, the associated gate counts are  still unfeasible on current day hardware platforms.
This leads to the situation that many-body wavefunctions of chemical systems, which are considered ``easy'' to describe within the classical algorithmic framework (e.g. using MP2 or CCSD), already require non-local and high-depth quantum circuits.
Ideally there should be a way to prepare these wavefunctions with low-depth and local quantum circuits where the circuit parameters can be determined robustly in a black-box fashion.\\
In this work, we will describe such an low-depth and local approach realized by physical principles and classical pre-computation resulting in optimized circuits through a separable pair approximation (SPA). In a way this can be seen as a \textit{hardware-efficient ansatz}~\cite{kandala2017hardware} based on physical principles, therefore combining the advantages of \textit{hardware-efficient} and \textit{physically inspired} approaches.\\

As an explicit embodiment, we will combine this ansatz with the directly determined pair-natural orbitals representation of~\cite{kottmann2020reducing} resulting in reduced gate counts and depths by several orders of magnitude - e.g. the BeH2 system from a circuit with 192 controlled-not gates~\cite{kottmann2020reducing} to a depth-5 circuit with 15 local controlled-not operations shown in Fig.~\ref{fig:beh2_circuit}.
Apart from low gate counts and depths, such wavefunctions prepared by the quantum circuits can be represented efficiently on a classical computer, which allows classical simulation and parameter optimization of the corresponding circuits.
The classically optimized quantum circuits could then be used to prepare initial states for more advanced quantum algorithms like the quantum phase estimation~\cite{aspuru2005simulated} as well as for variational algorithms that use this ansatz as a baseline for building more electron correlation. We will give explicit examples for both situations and identify good benchmark systems with small qubit sizes suitable for future improved schemes.
Our approach is integrated into the basis-set-free framework of Ref.~\cite{kottmann2020reducing} where we further improved the underlying surrogate model. We provide a full-stack open-source implementation in \textsc{tequila}~\cite{tequila}.

\section{Methodology}
Unitary coupled-cluster quantum circuits are formed from elementary $n$-electron excitation gates 
\begin{align}
    U_{\boldsymbol{pq}}\lr{\theta} = e^{-i\frac{\theta}{2} G_{\boldsymbol{pq}}}\label{eq:ucc_gate_general}
\end{align}
which describe excitations between spin orbitals $\boldsymbol{p}=\left\{p_0,p_1,\dots, p_{n}\right\}$ and $\boldsymbol{q}=\left\{q_0,q_1,\dots,q_{n}\right\}$ with $ p_k\neq q_l\; \forall (k,l)$ through the fermionic excitation generator
\begin{align}
    G_{\boldsymbol{pq}} = i\left(\prod_{k} a_{p_k}^\dagger a_{q_k} - h.c.\right)\label{eq:ucc_generator_general}.
\end{align}
The gate acts like a rotation on a subspace spanned by all configurations with spin-orbitals $\boldsymbol{p}$ occupied and $\boldsymbol{q}$ unoccupied and vice versa, while acting trivially (as the identity) on all other configurations
\begin{align}
    U_{\boldsymbol{p}\boldsymbol{q}}\lr{\theta} =& \cos\lr{\frac{\theta}{2}} -i\sin\lr{\frac{\theta}{2}}G_{\boldsymbol{pq}}\nonumber\\ &+ \lr{1-\cos\lr{\frac{\theta}{2}}}P^0_{\boldsymbol{pq}}.
\end{align}
Here, $P^0_{\boldsymbol{pq}}$ is the nullspace projector~\cite{kottmann2020feasible} of the generator $G_{\boldsymbol{pq}}$
\begin{align}
    P^0_{\boldsymbol{pq}} = 1 - \prod a^\dagger_{p_k}a_{p_k}a_{q_k}a^\dagger_{q_k} - \prod a^\dagger_{q_k}a_{q_k}a_{p_k}a^\dagger_{p_k}
\end{align}
and the generators $G_{\boldsymbol{pq}},P^0_{\boldsymbol{pq}}$ can be mapped to qubits via various transformations.\\ One of the most established qubit encodings, the Jordan-Wigner transformation, maps the creation and annihilation operators directly to qubit raising and lowering operators $\sigma^\pm = \frac{1}{2}\lr{\sigma_x+i\sigma_y}$ and $\sigma_z$ operators that ensure the correct anti-commutation properties 
\begin{align}
    a^\dagger_p \xrightarrow[Wigner]{Jordan}\sigma^+_p \prod_{k<p} \sigma^z_k.
\end{align}
Within the Jordan-Wigner encoding, each spin-orbital is mapped directly to a qubit. In terms of spatial orbitals this means we need twice as many qubits as spatial orbitals $\Nq=2\No$. In this work, we will use the Jordan-Wigner representation due to its intuitive simplicity. In future applications, we could imagine local encodings~\cite{chien2020custom, setia2018bravyikitaevsuperfast, derby2020compact} to be advantageous. In order to employ them within the methodology developed here, solely the bridging unitary (see Eq.~\eqref{eq:hcb_to_jw}) that maps from the hard-code Boson representation to the Jordan-Wigner representation needs to be adapted. 

\begin{table}[]
    \centering
\begin{tabular}{lr|rr|rr|rr}
\toprule
\multicolumn{2}{c}{ }&\multicolumn{6}{c}{Optimization level}\\
\multicolumn{2}{c}{ }&\multicolumn{2}{c}{2} &\multicolumn{2}{c}{1} &\multicolumn{2}{c}{0} \\
\midrule
\multicolumn{2}{l}{Method\phantom{aa}$N_\text{param}$}& \multicolumn{2}{c}{Depth\phantom{.}$N_\text{cnot}$}& \multicolumn{2}{c}{Depth\phantom{.}$N_\text{cnot}$} & \multicolumn{2}{c}{Depth\phantom{.}$N_\text{cnot}$}\\
\midrule
SPA        &     3 &     3 &     9 &    23 &    42 &    73 &   144\\
UpCCD      &     9 &    31 &    33 &    89 &   126 &   287 &   432\\
UpCCSD     &    27 &   548 &   465 &   606 &   558 &   804 &   864\\
UpCCGASD   &    45 &   122 &   177 &   219 &   330 &   564 &   840\\
2-UpCCGASD &    90 &   340 &   507 &   437 &   660 &  1124 &  1680\\
UpCCGSD    &    45 &   669 &   617 &   766 &   770 &  1111 &  1280\\
2-UpCCGSD  &    90 &  1434 &  1387 &  1531 &  1540 &  2217 &  2560\\
\bottomrule
\end{tabular}
    \caption{Details about the circuits used in Fig.~\ref{fig:results_beh2_bh3_n2} for the N2(6,12) and BH$_3$(6,12) systems with different levels of optimization: None (0), gate decompositions similar to Ref.~\cite{yordanov2020efficient} (1), including initialization in the HCB subspace according to Eq.~\eqref{eq:upccgsd} using SPA as initial part of the circuit (2) . UpCCGASD and similar circuits employ approximate singles (similar to ``A'' gates in ~\cite{gard2020efficient}) where $\sigma_z$ terms in the generators are neglected. }
    \label{tab:details_bh3}
\end{table}

\subsection{Paired Coupled-Cluster and the Hard-Core Boson Model}
In this section, we will describe and unify ideas from Refs.~\cite{elfving2020simulating, khamoshi2020correlating} that investigated paired unitary models with general unitary coupled-cluster approaches, in particular with the $k$-UpCCGSD~\cite{lee2018generalized} hierarchy. The combination of those approaches alone already leads to significantly reduced gate counts and depths which is further reduced by optimized gate decompositions from Refs.~\cite{yordanov2020efficient} (and similar in Ref.~\cite{anselmetti2021local}) and~\cite{gard2020efficient}. In the next section we will construct the separated pair ansatz and integrate it in the optimized $k$-UpCCGSD hierarchy. This will define a local, low-depth and classically simulable class of quantum circuits.\\

Paired unitary coupled-cluster models build their wavefunctions from single excitations and double excitations restricted to paired electrons in the same spatial orbitals $p$ and $q$.
The corresponding generator for a primitive unitary excitation operator is
\begin{align}
    G_{p_\uparrow p_\downarrow q_\uparrow q_\downarrow} \equiv \tilde{G}_{pq} = i\lr{a_{p_\uparrow}^\dagger a_{q_\uparrow} a_{p_\downarrow}^\dagger a_{q_\downarrow} - h.c}.\label{eq:ucc_paired_generators}
\end{align}
In the Jordan-Wigner encoding, all $\sigma_z$ operations in these restricted double excitations cancel out, yielding a qubit excitation generator
\begin{align}
    \tilde{G}_{pq} \xrightarrow[\text{Wigner}]{\text{Jordan-}} \tilde{G}^\text{JW}_{pq} = i\left(\sigma^+_{p_\uparrow} \sigma_{q_\uparrow}^- \sigma^+_{p_\downarrow} \sigma_{q_\downarrow}^- - h.c.\right)\label{eq:qubit_excitation_generator}
\end{align}
that can be compiled into a unitary circuit with depth 22 and 13 CNOT gates~\cite{yordanov2020efficient} (see Ref.~\cite{anselmetti2021local} for an alternative construction).
If we restrict the ansatz to only paired doubles, \textit{e.g.} UpCCD, and start from a restricted reference wavefunction, the resulting wavefunction will consist solely of doubly occupied configurations. Instead of representing the spin-up and spin-down part of a spatial orbital individually with two qubits, we can now encode the doubly occupied or non-occupied spatial orbitals by a single qubit. This restriction is commonly referred to as a hard-core Boson (HCB) model of the original fermionic system and has been employed for variational quantum algorithms in reduced qubit representations~\cite{elfving2020simulating, khamoshi2020correlating}.\footnote{Instead of hard-core Boson, the model is often also labelled as seniority-zero or simply as a doubly-occupied or paired model.The term seniority-zero results from the seniority quantum number of the associated wavefunction which is a quantifier for unpaired electrons in the system.} Associated classical algorithms are paired coupled-cluster (pCCD) or doubly-occupied configuration interaction (DOCI).\\

If the wavefunction is restricted to paired excitations, the whole wavefunction can be represented in the hard-core Boson representation by encoding the pair-excitation generators of Eq.~\eqref{eq:ucc_paired_generators} as
\begin{align}
    \tilde{G}_{pq} \xrightarrow[\text{Boson}]{\text{hard-core}} \tilde{G}^\text{HCB}_{pq}= i\lr{\sigma^+_p\sigma^-_q - h.c.}\label{eq:paired_double_excitation_generator_hcb}
\end{align}
where qubits $p,q$ represent electron pairs in spatial orbitals $p,q$. Thus $\No$ spatial orbitals are mapped to $\No$ qubits. This is conceptually the same as in Ref.~\cite{elfving2020simulating} and the HCB-Hamiltonian can be mapped to qubits using the same principles.
Here, we aim to prepare good initial states of the original Hamiltonian, so we need to transfer the hard-core Boson wavefunction into a regular Jordan-Wigner represented qubit wavefunction. This can be achieved throughout a series of controlled-not operations
\begin{align}
    U_{\text{HCB}}^{\text{JW}} = \prod_{p=1}^{\No} \text{CNOT}\lr{p_\uparrow, p_\downarrow}\label{eq:hcb_to_jw}
\end{align}
that transfer the occupation information of the hard-core Boson wavefunction to a second register of qubits. The original qubit register $\boldsymbol{p}_\uparrow$, that represented electron pairs in the hard-core Boson representation, will then represent spin-up electrons in the Jordan-Wigner representation and the additional register $\boldsymbol{p}_\downarrow$ will represent the corresponding spin-down electrons. 
An optimized UpCCGSD circuit is then constructed as
\begin{align}
    U_\text{pCCGSD}\lr{\boldsymbol{\theta^\text{S}, \boldsymbol{\theta^\text{D}}}} = U_\text{S}\lr{\boldsymbol{\theta^\text{S}}}U^{\text{JW}}_\text{HCB} U_D^\text{HCB}\lr{\boldsymbol{\theta^\text{D}}} U^\text{HCB}_\text{HF},\nonumber
\end{align}
\begin{align}
    U_D^\text{HCB}\lr{\boldsymbol{\theta}} &= \prod_{p<q} e^{-i\frac{\theta_{pq}}{2} \tilde{G}_{pq}^\text{HCB} }\nonumber,\\
    U_S\lr{\boldsymbol{\theta}} &= \prod_{p<q}  e^{-\frac{\theta_{pq}}{2}\left( G^\text{JW}_{p_\uparrow q_\uparrow} + G^\text{JW}_{p_\downarrow q_\downarrow}\right)}, \label{eq:upccsd}
\end{align}
representing the pair-double excitations in the hard-core Boson representation and the singles excitations in the standard Jordan-Wigner representation. The Hartree-Fock reference for an $\Ne$ electron system is initialized as $U^\text{HCB}_\text{HF} = \prod_{p\leq \N{e}/2} \sigma^x_p$.\\
Note, that this particular construction of UpCCGSD circuits partially fixes the order of the primitive circuits as it requires the singles block to be separated from the doubles block. This high-level ordering is however often empirically preferred~\cite{grimsley2019trotterized}.\\
Further integration into the $k$-UpCCGSD~\cite{lee2018generalized} hierarchy can be obtained by adding further layers of excitations
\begin{align}
    U^{\left(k\right)}_\text{pCCGSD} = \prod_{l=2}^{k}\lr{ U_\text{S}\lr{\boldsymbol{\theta}^\text{S}_l}U_\text{D}\lr{\boldsymbol{\theta}^\text{D}_l} }U_\text{pCCGSD}\lr{\boldsymbol{\theta}^\text{S}_1,\boldsymbol{\theta}^\text{D}_1}\label{eq:upccgsd}
\end{align}
with $U_\text{D}$ constructed similar to $U_\text{D}^\text{HCB}$ by using $\tilde{G}^\text{JW}_{pq}$.

\begin{table}[]
    \centering
    \begin{tabular}{lcccccccc}
    \toprule
Molecule($\Ne$,$\Nq$) & $N_\text{param}$& $N_\text{cnot}$ & Depth \\ 
\midrule
H$_2$(2,4)            &     1 &     3 &     3 \\
LiH(2,10)             &     4 &    15 &    18 \\
BeH$_2$(4,8)          &     2 &     6 &     3 \\
BeH$_2$(6,14)         &     4 &    15 &     7 \\
BH$_3$(6,12)          &     3 &     9 &     3 \\
N$_2$(6,12)           &     3 &     9 &     3 \\
C$_2$H$_4$(12,24)     &     6 &    18 &     3 \\
H$_2$O$_2$(14,28)     &     7 &    21 &     3 \\
C$_2$H$_6$(14,28)     &     7 &    21 &     3 \\
C$_2$H$_6$(2,12)      &     5 &    19 &    23 \\
C$_2$H$_6$(14,84)     &     35&   133 &    23 \\
\bottomrule
    \end{tabular}
    \caption{Resource requirements for SPA circuits according to Eq.~\eqref{eq:pno_upccd} used in this work. We show minimal configurations ($N_\text{q}=2\cdot N_\text{e}$ with two spatial orbital for each electron pair) with low-depth circuits as well as active space configurations with more spatial orbitals for individual electron pairs. E.g. LiH(2,10) and C$_2$H$_6$(2,12) with 5 and 6 spatial orbitals for a single electron pair and C$_2$H$_6$(14,84) with 6 spatial orbitals for each active electron pair. The BeH$_2$(6,14) circuit with two spatial orbitals for both active electron pairs and a single spatial orbital for the core orbital is shown explicitly in Fig.~\ref{fig:beh2_circuit}}
    \label{tab:gate_counts}
\end{table}

\begin{figure}
    \centering
    \begin{tabular}{ccc}
    \includegraphics[width=0.45\textwidth]{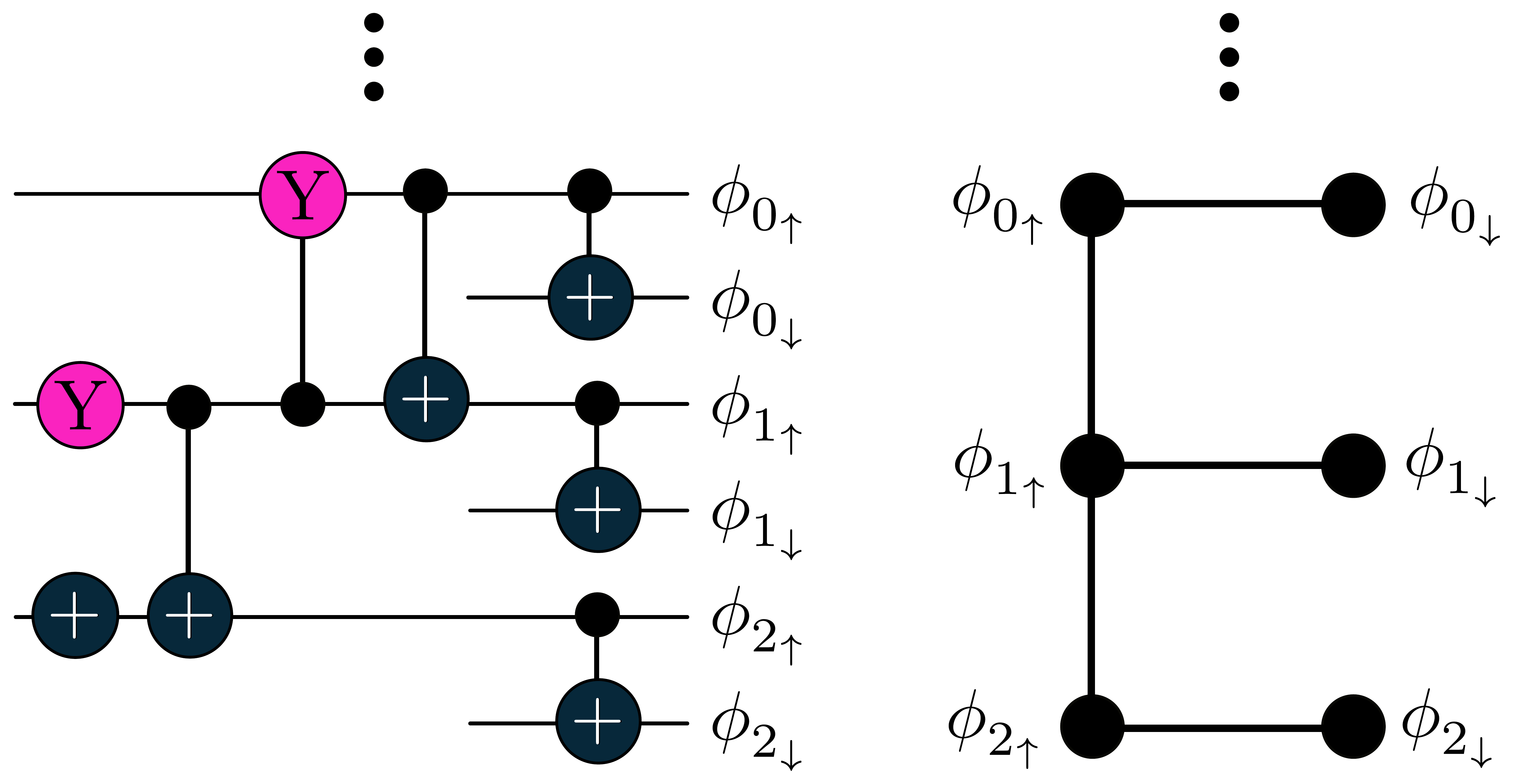}
    \\
    \end{tabular}
    \caption{Template SPA circuit (left) and required qubit connectivity (right) for a single electron pair represented by 3 spatial orbitals in ladder arrangement. The first \textit{NOT} operation on the lower left prepares the two-electron reference state in the hard-core Boson representation ($U_\text{HF}^\text{HCB}$ in Eq.~\eqref{eq:pno_upccd}). The \textit{CNOT} layer on the right transfers the doubly-occupied wavefunction in the hard-core Boson representation to the Jordan-Wigner representation ($U_\text{HCB}^\text{JW}$ in Eq.~\eqref{eq:pno_upccd}).}
    \label{fig:circuit_compiling_cartoon}
\end{figure}

\begin{figure*}
    \centering
    \includegraphics[width=0.8\textwidth]{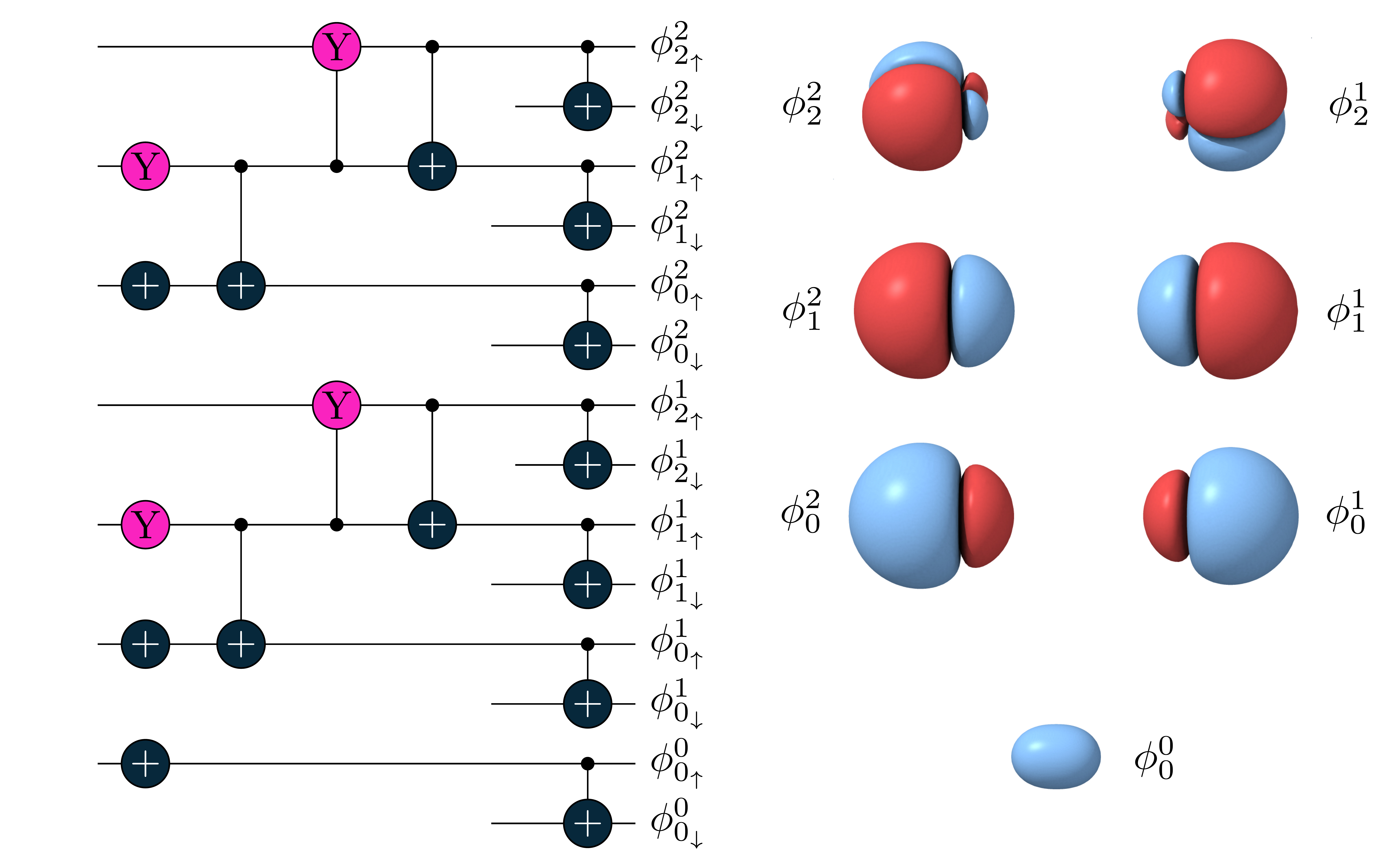}
    \caption{Example: Directly compiled low-depth SPA ladder arrangement for the BeH$_2$ molecule initializing a wavefunction as in Eq.~\eqref{eq:paired_wfn}. Controlled $R_y$ rotations can be compiled into two controlled-not operations and three single qubit rotations, leading to an overall CNOT count of 15 and a circuit depth of 7. The circuit corresponds to the BeH$_2$(6,14) circuit in Tab.~\ref{tab:gate_counts}. Pink gates represent individually parametrized Pauli-Y rotations $R_y(\theta_k)=e^{-i\frac{\theta_k}{2} \sigma_y }$ and $+$ labels represent (controlled)-not operations. }
    \label{fig:beh2_circuit}
\end{figure*}

\subsection{Separable Pair Ansatz (SPA)}
In the following, we will describe a separable pair approximation in a general framework and combine it with the circuit compilation strategies of the last section to construct a classically tractable circuit class with significantly reduced gate count. \\

Assume we have an $\Ne$ electron system and we want to create a wavefunction of $\frac{\Ne}{2}$ electron pairs. This separable pair (SP) wavefunction can be written as 
\begin{align}
    \ket{\pairwfn} = \bigotimes_{k=1}^{\Ne/2} \ket{\psi_k}\label{eq:paired_wfn}
\end{align}
where $\ket{\psi_k}$ are electron pair functions, that can themselves be represented by a linear combination of tensor-products of $|{S}_k|$ one-electron functions (spin-orbitals)
\begin{align}
    \ket{\psi_k} = \sum_{mn} c^k_{mn} \ket{\phi^k_m}\otimes\ket{\phi^k_n}.
\end{align}
Each pair-function $\ket{\psi_k}$ is represented by an individual set of orbitals ${S}_{k} = \left\{\ket{\phi^k_l}, l=0,\dots,|{S}_k|-1\right\}$ and we will furthermore require all orbitals to be orthonormal
\begin{align}
   \braket{\phi^k_l}{\phi^{k'}_{l'}} = \delta_{kk'}\delta_{ll'}.
\end{align}
In order to generate the wavefunction $\ket{\pairwfn}$ in a qubit representation we only require the unitaries $U_k$ that create the pair-functions $\ket{\psi_k}$.
One strategy to realize $U_k$ is through one- and two-electron excitation gates as in Eq.~\eqref{eq:ucc_gate_general} acting on a closed-shell initial state
\begin{align}
    \ket{\pairwfn} = \prod_k^{\Ne/2} U_{k}\left(\boldsymbol{\theta}_k\right)U_\text{HF} \ket{00\dots0}\label{eq:product_structure}.
\end{align}
Note that the PNO-UpCCGSD circuits of Ref.~\cite{kottmann2020reducing} have exactly this product structure and  the resulting wavefunctions can be fully simulated classically without facing an exponential memory bottleneck. It is only required to store $\frac{\Ne}{2}$ individual pair functions that can each be fully described with $\mathcal{O}\left(|{S}_k|^2\right)$ coefficients. 
The classical optimization results directly in optimized parameters of a low-depth quantum circuit that is ready to be used within an extended quantum algorithm. Furthermore, such classically simulable circuits naturally define a minimum benchmark that variational quantum algorithms need to match in order to be considered for a potential advantage.\\
Note, that the associated variational algorithm will minimize the expectation value of the parametrized product of pair-functions over the full electronic Hamiltonian
\begin{align}
    E = \min_{\boldsymbol{\theta}}\expval{H}{\pairwfn\left(\boldsymbol{\theta}\right)}.\label{eq:pair-vqe}
\end{align}
In other words, while the wavefunction has a product structure, the individual pair-functions are not independent but coupled through the Hamiltonian. The latter basically defines a mean-field model for pairs, similar to generalized valence bond models (GVB) with strong orthogonality condition~\cite{larsson2020minimal}.\\

The individual pair-functions can again be restricted to be occupied by hard-core Bosons only (\textit{i.e.} each spatial orbital can only be occuied by spin-paired electrons). This allows the same reduction in qubits as described in the previous section where each orbital is now represented by a qubit. Circuits can now be constructed by arranging double excitations (as in $U_\text{D}^\text{HCB}$ in Eq.~\eqref{eq:upccsd}) in the hard-core Boson representation in order to construct the individual pair function unitaries
\begin{align}
    U_\text{SPA} = U_\text{HCB}^\text{JW} \lr{\prod_k^{\Ne/2} \prod_{l\in {S}_{k}} e^{-i\frac{\theta^k_{l}}{2}\tilde{G}^\text{HCB}_{l,l+1}} } U_\text{HF}^\text{HCB}.\label{eq:pno_upccd}
\end{align}
At this point, the memory requirements to represent the associated wavefunction are further reduced to $\mathcal{O}\left(|{S}_k|\right)$ for each pair.
In Eq~\eqref{eq:pno_upccd} we chose a ladder arrangement of double excitations within each pair that requires only local connectivity of the associated qubits.
Since the unitaries that prepare the individual pair-functions act on an initial product state prepared by $U_\text{HF}^\text{HCB}$, they can be efficiently compiled into controlled-not and controlled rotation gates as
\begin{align}
    e^{-i\frac{\theta^k_{l}}{2}\tilde{G}^\text{HCB}_{l,l+1}}\rightarrow \text{CR}_y(\theta^k_l,{S}_{k}^l,{S}_{k}^{l+1}) \cdot\text{CNOT}({S}_k^{l+1},{S}_k^l)
\end{align}
where the control on the CR$_y$ can be dropped for the very first instance. This circuit construction procedure is schematically depicted in Fig.~\ref{fig:circuit_compiling_cartoon}.
Alternatively the doubles can be arranged canonically by exciting from the ``reference orbital'' $S_k^1$ to all other orbitals in ${S}_k$ like in an UpCCD ansatz for the individual pair.\\
Both approaches have the same expressibility but differ in their locality and in their behaviour under optimization where the more local ladder arrangement usually requires more iterations in gradient based optimizations starting from an initial point with all angles in zero. An intuitive explanation is, that within the first $m$ iterations, all gradients except for the $\theta^k_l$ with $l\leq m$ iterations of each pair are naturally zero since the corresponding qubits are not occupied yet in the wavefunction as the occupation of the qubits gradually needs to be distributed from the first orbital to the last. Within the canonical arrangement all qubits can be occupied after the first iteration. A more natural starting point for the ladder approximation would be a finite value for all angles. In the specific embodiment of this work, the orbitals that represent the individual pairs are ordered through the classical surrogate model that determines them (see next section). From this property we can already assume that the initial values for the angles $\theta_l^k$ should decrease in magnitude with growing $l$. The exact behaviour and sophisticated initialization of the correct signs of the angles could be an interesting testing case for currently emerging initialization protocols.~\cite{cervera2020meta, sauvage2021flip}

\subsection{Orbital Determination}
\begin{figure*}
    \centering
    \begin{tabular}{cc}
    \includegraphics[width=0.4\textwidth]{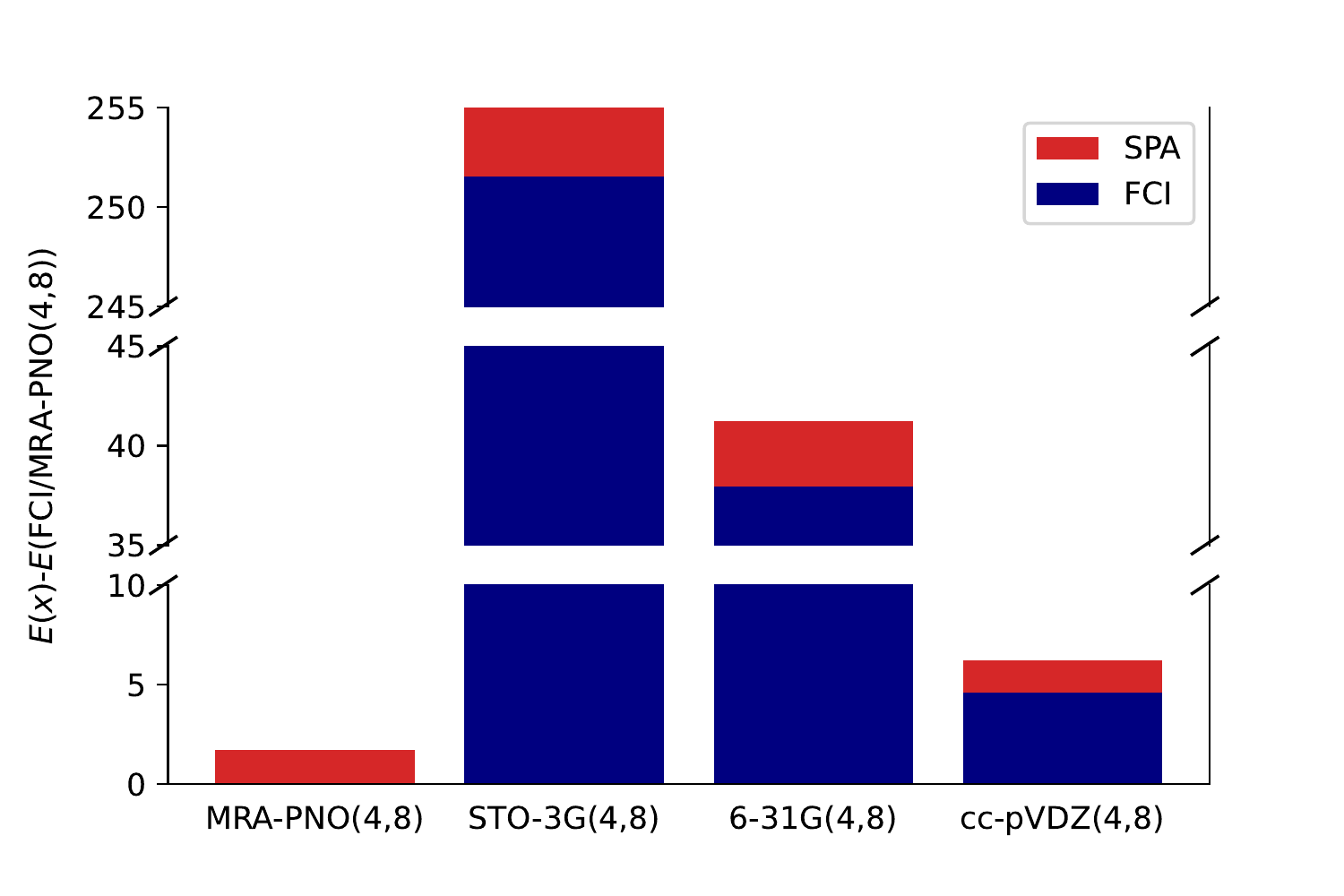}
    \shiftleft{6cm}{\raisebox{3.75cm}[0cm][0cm]{(a)}}
    &
    \includegraphics[width=0.4\textwidth]{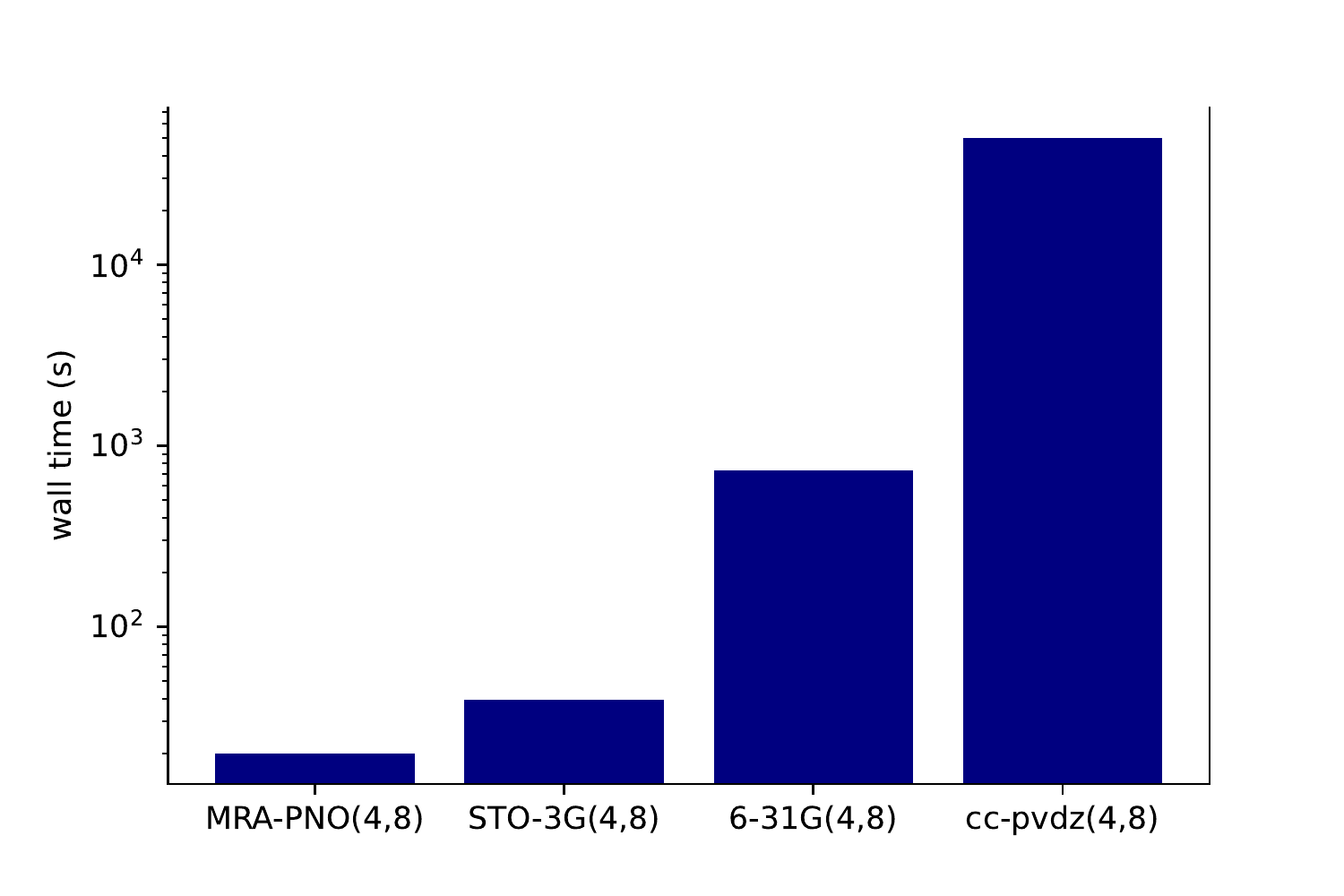}
    \shiftleft{6cm}{\raisebox{3.75cm}[0cm][0cm]{(b)}}
    \\
    \end{tabular}
    \caption{Orbital determination of a minimal SPA ansatz for BeH$_2$(4,8) with bond distance 1.5\AA. Energy differences w.r.t FCI/MRA-PNO(4,8) (left) and total simulation walltime (right). The SPA circuit corresponds to the sub-circuit of Fig.~\ref{fig:beh2_circuit} including orbitals $\left\{ \phi_0^1, \phi_1^1, \phi_0^2, \phi_1^2 \right\}$. For MRA-PNO(4,8) those orbitals correspond to the ones shown in Fig.~\ref{fig:beh2_circuit}, for the Gaussian basis sets (STO-3G, 6-31G, cc-pVDZ) they where optimized through the standard orbital optimization procedure implemented in \textsc{pyscf}\cite{pyscf2}. Here, the best four spatial orbitals for the SPA circuit represented by the corresponding orbitals of the basis set are determined. Further details are given in the appendix.}
    \label{fig:spa_mra_vs_gbs}
\end{figure*}

In the previous section we constructed optimized quantum circuits that prepare products of electron pairs with each of the pairs represented by independent orbital sets ${S}$.
In order to determine these orbitals we will resort to a modified approach of Ref.~\cite{kottmann2020reducing} that takes a classical, basis-set-independent surrogate model~\cite{kottmann2020direct}. Alternatively the orbitals can be (further) optimized through standard orbital optimization techniques.
In the following we will briefly describe the two techniques.

\paragraph{Arbitrary Orbital Source:}

In cases where the assignment of the given orbitals to the sets ${S}_k$ is unclear, we propose to employ standard orbital optimization: 1. Take a set of orthogonal orbitals. 2. Distribute them into the orbital sets ${S}_k$. 3. Construct the SPA circuits according to Fig.~\ref{fig:circuit_compiling_cartoon}. 4. Optimize the SPA angles. 5. Optimize the orbitals for the current SPA wavefunction. Repeat steps 3-5 until overall convergence.
Standard approaches to optimize the orbitals are through the first order expanded expectation value of the orbital-rotated Hamiltonian \textit{w.r.t} the given wavefunction. In Fig.~\ref{fig:spa_mra_vs_gbs} we demonstrated this on a small example using the implementation ~\cite{pyscf2} of~\cite{sun2017coiterative} where we provide more details in the appendix. For an introduction on orbital-optimization in the context of variational quantum algorithms we refer to Refs.~\cite{sokolov2020quantum, mizukami2020, yalouz2021stateaveraged}.\\

\paragraph{MRA-PNOs}
In the implementation of Ref.~\cite{kottmann2020direct} the pair-natural orbitals (PNOs) for the MP2 model are determined directly on an adaptive grid via a multiresolution analysis (MRA). The algorithm determines the occupied Hartree-Fock orbitals $\phi_k$ ($k\leq \frac{\Ne}{2}$) and pair-natural orbitals $\phi_{kl}^{r}$ ($k\leq l\leq\frac{\Ne}{2}, r\leq\textsc{maxrank}$). We will briefly refer to both types of orbitals as MRA-PNOs. In the previous work~\cite{kottmann2020reducing} MRA-PNOs were employed to construct system-adapted qubit Hamiltonians and initial attempts of designing unitary coupled-cluster circuits that exploit the structure of the orbitals were formulated.  
The here proposed separable pair approximation generalizes these attempts.\\
We adapted the implementation of Ref.~\cite{kottmann2020direct} in order to compute only the ``diagonal'' MP2-PNOs $\phi_{kk}^r$. After determining the Hartree-Fock orbitals $\phi_k\equiv\phi_k^0$ and the associated PNOs $\phi_{kk}^{r}\equiv\phi_{k}^{r}$ we perform a global symmetric orthogonalization (diagonalization of the overlap matrix) on all PNOs and assign them to the pair sets ${S}_k=\left\{ \phi_k^0, \tilde{\phi}^1_k, \dots \tilde{\phi}_k^\textsc{maxrank} \right\}$ where $\tilde{\phi}^r_k$ denote the orthonormalized PNOs.
The runtime of the original MRA-PNO implementation scales as $\mathcal{O}\lr{N_\text{pairs} \Ne \textsc{maxrank}^2}$ where the factor $\Ne$ comes from the exchange operator in the MP2-PNO equations.
As we only compute ``diagonal'' pairs here, we have $\N{pairs}=\frac{\Ne}{2}$ leading to an overall quadratic dependence $\mathcal{O}\lr{\Ne^2}$ on system size for the orbital determination via MRA-PNOs. Not that the parameter \textsc{maxrank} does not depend on the system size as it was shown in~\cite{kottmann2020direct}.
It was shown before, that this cost can be mitigated to near-linear behaviour by efficiently exploiting locality in the multiresolution representation.~\cite{yanai2004exchange} In the current implementation this is however not included yet.\\

In Fig.~\ref{fig:spa_mra_vs_gbs} we compare SPA with MRA-PNOs and optimized Gaussian basis sets on an explicit example illustrating the advantages of MRA-PNOs in accuracy and runtime. Here we note, that initial PNOs could also be computed in Gaussian basis sets (see in this context Refs.~\cite{gonthier2020identifying, deprince2013accurate}) and adapted for qubit Hamiltonians similar as in Ref.~\cite{kottmann2020reducing}. We expect that this would decrease the runtimes of Fig.~\ref{fig:spa_mra_vs_gbs} but not the accuracy that is expected to further converge towards the SPA/MRA-PNO(4,8) result with increasing basis size.

\begin{figure*}
    \centering
    \begin{tabular}{ccc}
    (a) & (b) & (c)\\
    \includegraphics[width=0.32\textwidth]{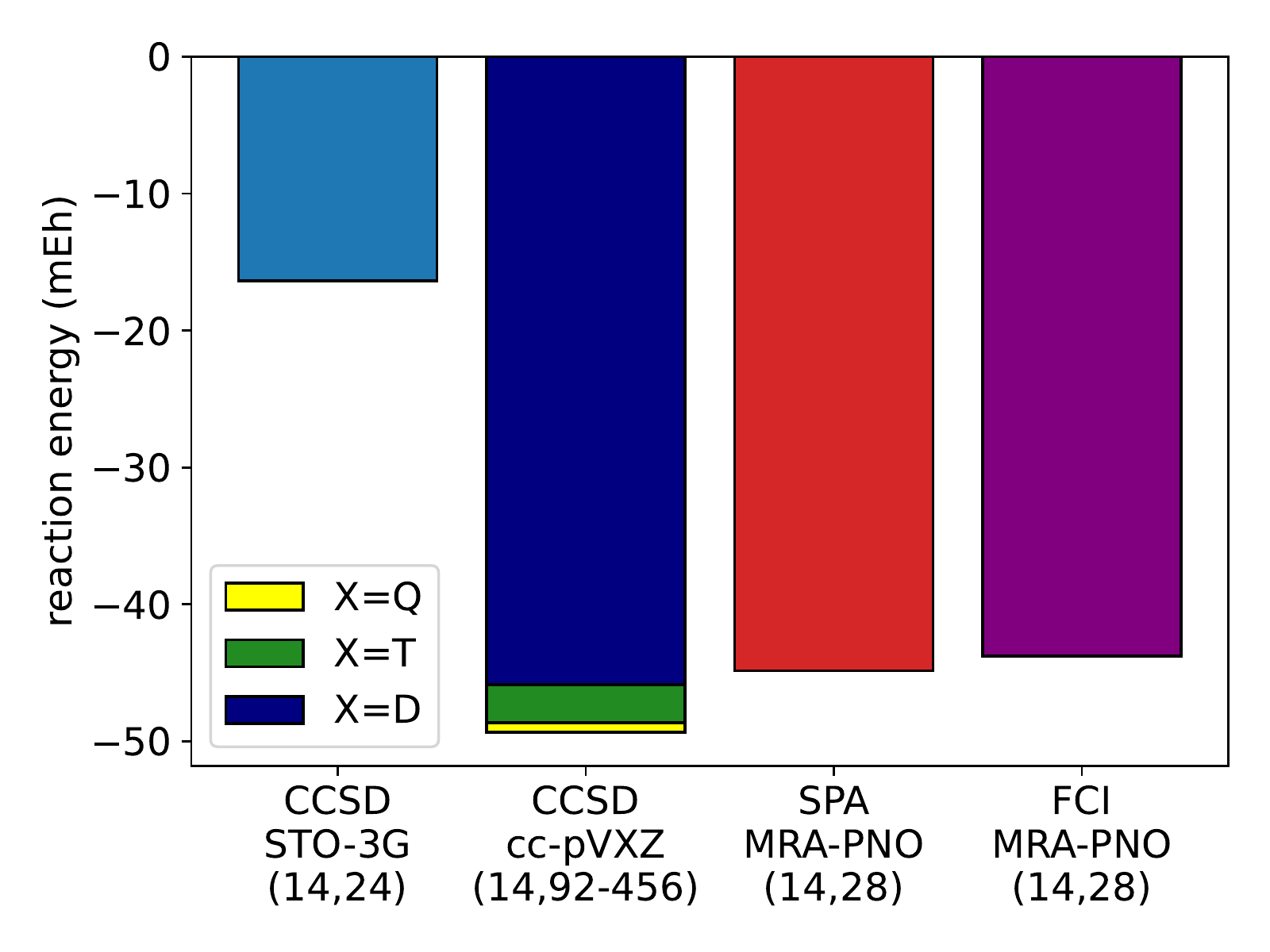}&
    \includegraphics[width=0.32\textwidth]{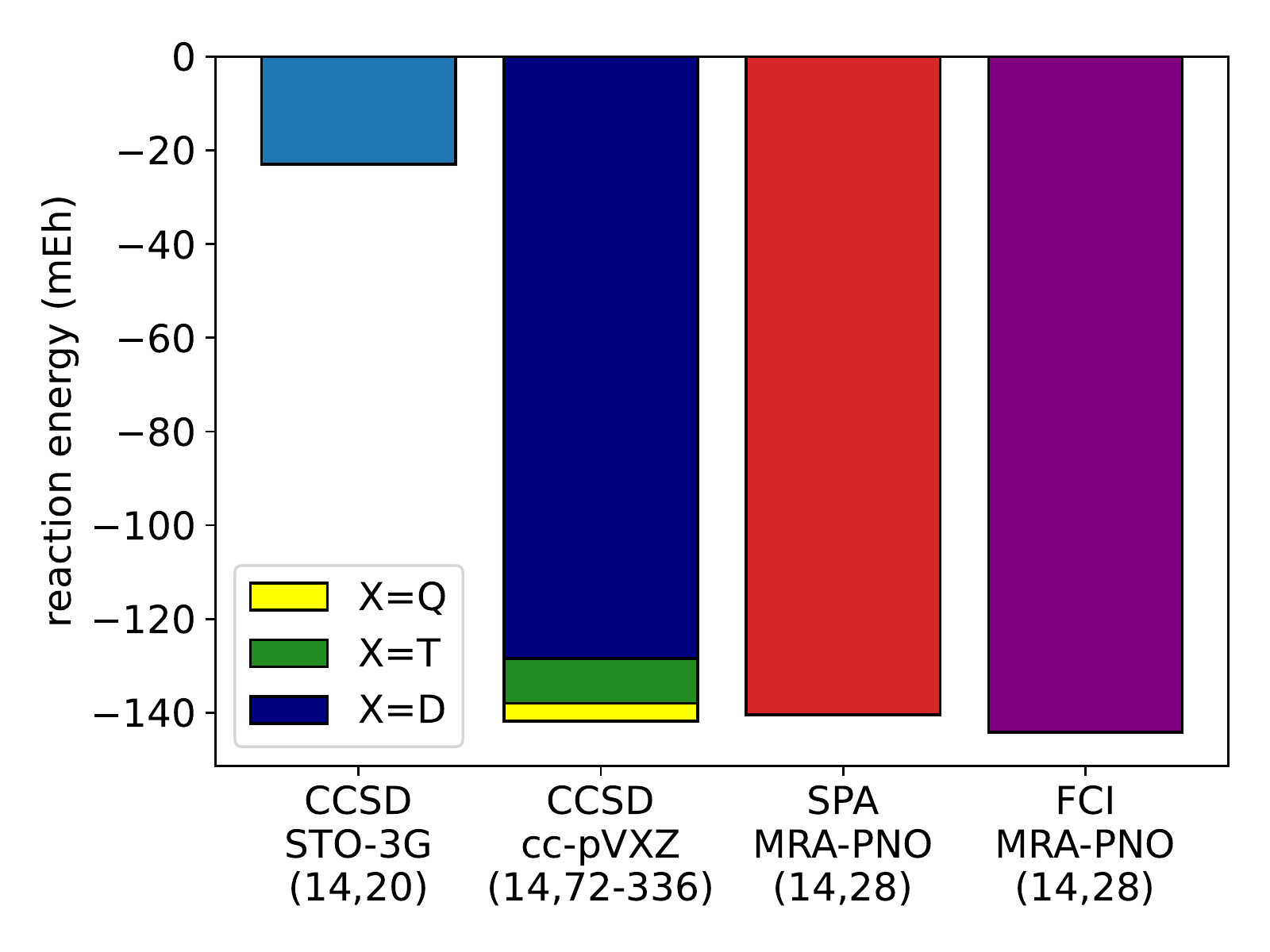}&
    \includegraphics[width=0.32\textwidth]{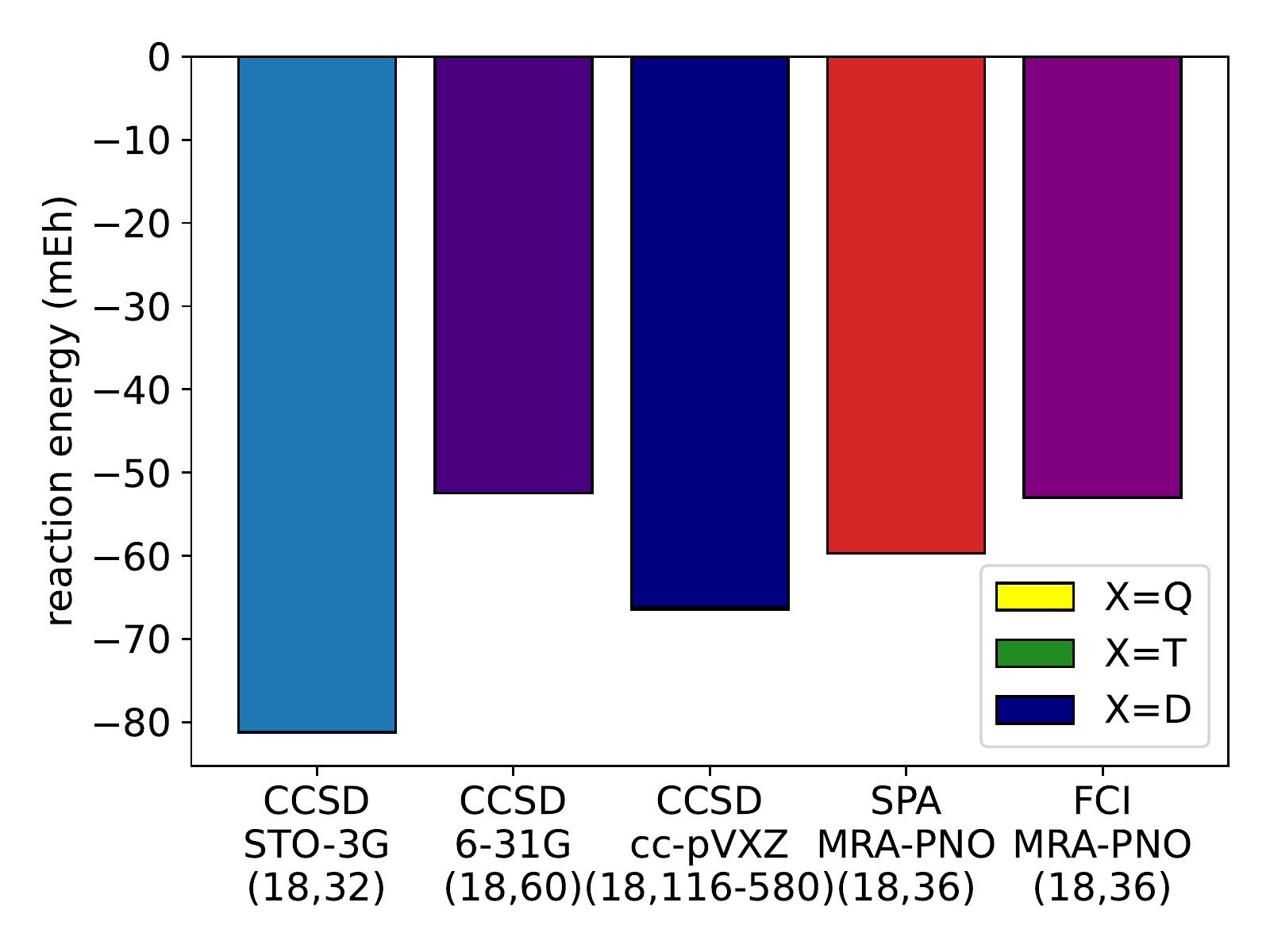}\\
    \end{tabular}
    \caption{Example Reactions: Performance of the SPA model in a system adapted orbital basis determined by MRA-PNOs, for chemical reaction energies ($E_\text{products} - E_\text{educts}$) compared to standard basis sets. Results are labeled as {method/basis/($N_\text{e}$,$N_\text{q}$)}. Reactions are CH$_3$OH + $\text{H}_{2}$ $\rightarrow$ CH$_4$ + H$_2$O (a) H$_2$O$_2$ + $\text{H}_{2}$ $\rightarrow$ $2\cdot$ H$_2$O (b) C$_2$H$_4$ + $\text{H}_{2}$ $\rightarrow$ C$_2$H$_6$ (c)}
    \label{fig:reactions}
\end{figure*}

\begin{figure*}
    \centering
    \includegraphics[width=1.0\textwidth]{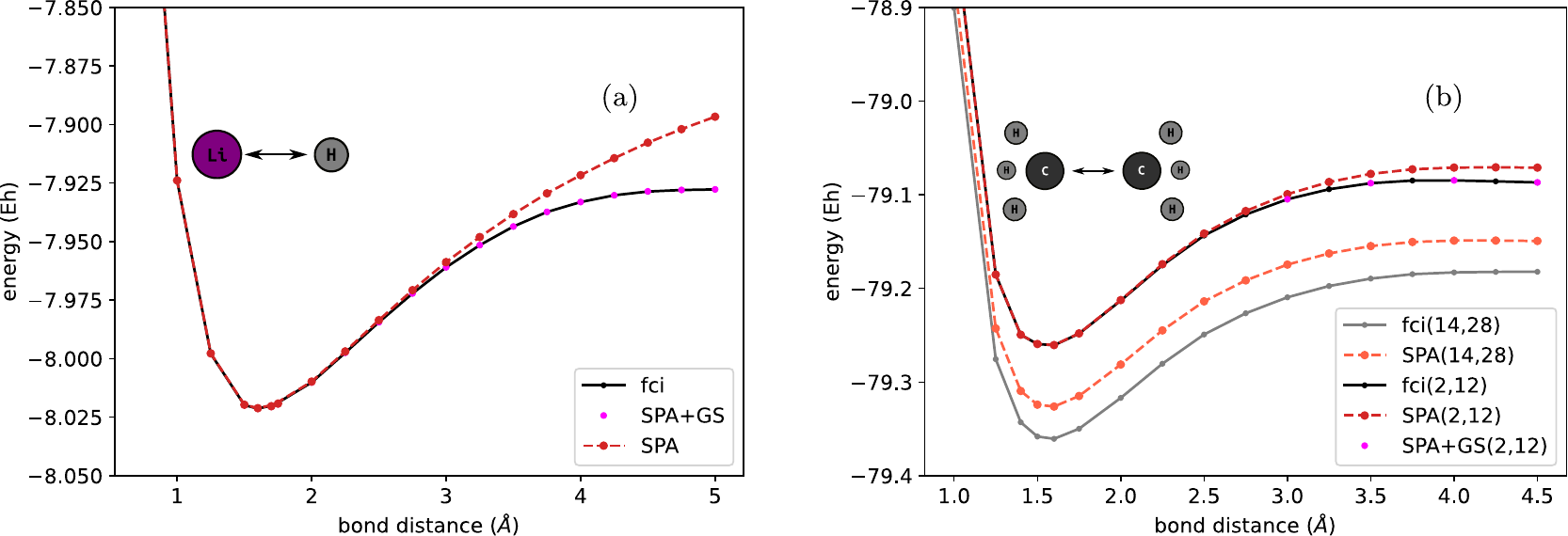}
    \caption{Simple Model Systems: Single bond dissociation of LiH(2,12) (a) and C$_2$H$_6$(2,12) (b). SPA circuits are constructed as in Eq.~\eqref{eq:pno_upccd} and a single layer of generalized orbital rotations is added. The performance of SPA+GS is in this case equivalent to orbital optimized SPA with orbital optimization similar to Refs.~\cite{mizukami2020, sokolov2020quantum,yalouz2021stateaveraged}. In addition we show C$_2$H$_6$(14,28) with all active electron pairs represented by two spatial orbitals (same representation as in Fig.~\ref{fig:reactions}). The associated non-parallelity errors (difference between largest and smallest absolute error) with FCI/MRA(14,28) as reference are: 3 for SPA/MRA(14,28), 18 for FCI/MRA(2,12) and 25 for SPA/MRA(2,12) millihartree.}
    \label{fig:lih_and_c2h6}
\end{figure*}

\begin{figure*}
    \centering
    \includegraphics[width=1.0\textwidth]{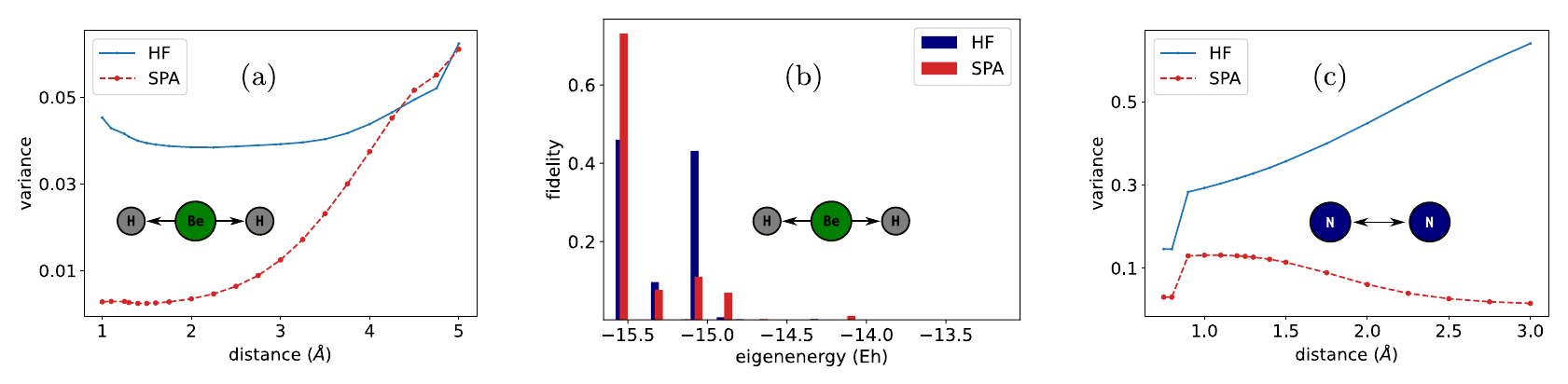}
    \caption{Variances $\|\expval{H}_U^2 - \expval{H^2}_U\|$ of HF (best classical mean-field solution) and SPA models for BeH$_2$(4,8) (a) and N$_2$(6,12) (c) as a quantifier for closeness to eigenstates. Shown at the center (b) are the fidelities $\|\braket{\Psi_U}{\Psi_\text{exact}}\|^2$ with respect to all eigenstates of the BeH2$_2$(4,8) with both Be-H bond distances at 5.0~{\AA}ngstrom, providing more details on the eigenstates that overlap with the trial wavefunction. }
    \label{fig:results_variances}
\end{figure*}

\section{Applications}
In the following, we will demonstrate explicit use-cases of the SPA and illustrate how it can potentially be used as initial state for other quantum algorithms. As in~\cite{kottmann2020reducing} we denote molecular representations as ``Molecule-Name($\Ne$,$\Nq$)'' with number of electrons $\Ne$ and number of qubits $\Nq=2\No$ necessary in a direct mapping of the spin orbitals.
In Fig.~\ref{fig:lih_and_c2h6} we compute single bond stretches of the LiH and C$_2$H$_6$ molecules, similar to Ref.~\cite{kottmann2020reducing}. As expected, the separable pair ansatz performs well at not too far stretched bond distances. The shortcoming can be overcome by including orbital rotations in the form of a generalized singles layer in the circuit. This performance can be expected to be equivalent to an orbital-optimized form of SPA as it was used in Fig.~\ref{fig:spa_mra_vs_gbs} for standard basis sets.
We also included a calculation with a larger active space C$_2$H$_6$(14,28) to confirm that the performance of the separable pair ansatz stays consistent. Here we observe a consistent energy difference to the FCI energies (non-parallelity error of 3 millihartree) resulting from the missing correlation between the pairs. Although this does not represent a rigorous benchmark, we anticipate that the separable pair ansatz will be an appropriate model for single bond reactions and organic equilibrium structures, especially in an orbital optimized extension similar to related classical methods like pCCD~\cite{henderson2014}, its orbital-optimized variants~\cite{scuseria1987optimization,bozkaya2011quadratically, kossoski2021excited}, and low-order matrix-product states respectively generalized valence bond models~\cite{larsson2020minimal}. \\

\begin{figure*}
    \centering
    \includegraphics[width=1.0\textwidth]{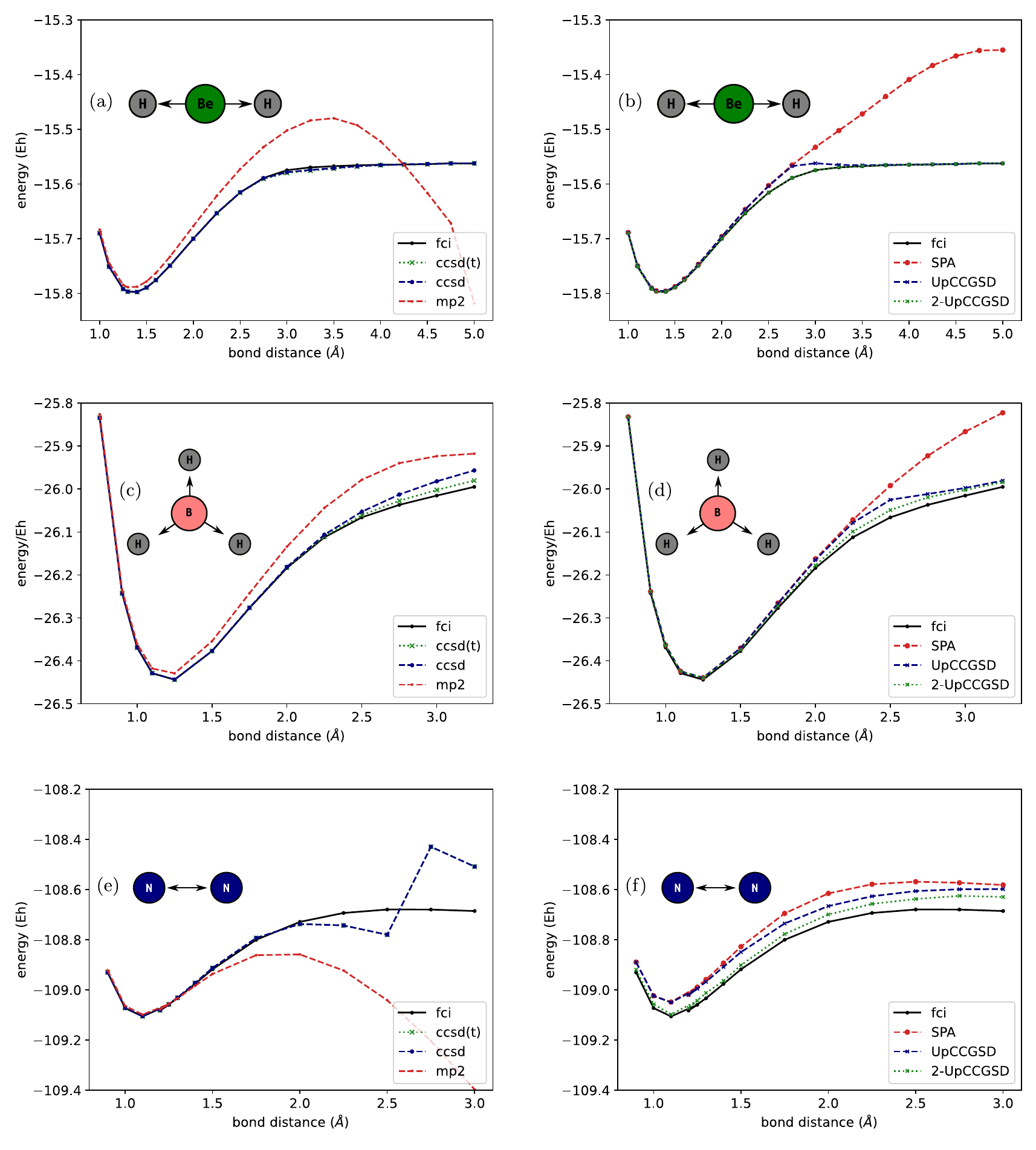}
    \caption{Challenging Model Systems: Comparison of standard classical methods (left - (a),(c),(e)) and pair-restricted quantum circuits (right - (b), (d), (f)) for the bonding electron pairs in BeH$_2$(4,8) (a)-(b), BH$_3$(6,12) (c)-(d) and N$_2$(6,12) (e)-(f). See Tab.~\ref{tab:details_bh3} for the required resources. }
    \label{fig:results_beh2_bh3_n2}
\end{figure*}

In Fig.~\ref{fig:results_beh2_bh3_n2} we computed double and triple bond dissociation of more challenging systems and compared classical methods with SPA and extensions in the $k$-UpCCGSD hierarchy. With BeH$_2$(4,8) and N$_2$(6,12) we included 8 and 12 qubit test systems which show variational breakdowns of standard methods  (MP2, CCSD and CCSD(T)) from classical quantum chemistry. All quantum models naturally don't show variational breakdowns. Furthermore we assume that the oscillating behaviour of CCSD is due to convergence problems. The SPA behaves fairly consistent over all three molecules, but, as for the single bond stretches, it shows large energy deviations for far stretched structures. In this case, including orbital rotations does improve, but not fully resolve, these differences.
The N$_2$(6,12) molecule remains the most challenging one; here, not even 2-UpCCGSD can reach FCI accuracy in all points. It was however shown before, that more layers of the ansatz systematically converge towards the FCI energy.~\cite{lee2018generalized}. We note the small basis deficiency of the MRA-PNOs at the stretched instances of N$_2$ which is inherited from the MP2 surrogate~\cite{kottmann2020reducing}.\\

An intuitive explanation for the behaviour of the SPA on the so far discussed examples is that chemical bonds are modelled as individual electron pairs - usually depicted as single lines in the graph representation of a molecule. The SPA wavefunction follows that chemical intuition. One interpretation is, that the SPA wavefunction provides an accurate description for instances of molecules where the graphical representation of the molecule is a good description. The chemical interpretation of the challenges arising with N$_2$ is here, that a triple bond is more than three single bonds. Consequently the quantum circuit to describe the $6$-electron system that forms this triple bond needs to go beyond separable pairs.

In Fig.~\ref{fig:results_variances} we performed a small numerical study where we assume to have access to a fault tolerant architecture, that is capable of performing a molecular quantum phase estimation~\cite{aspuru2005simulated}. In this proposed algorithm the physical measurement process of the full electronic Hamiltonian is implemented, and for simplicity we will assume a numerically exact implementation.
If a trial state $\ket{\Psi} = \sum_k c_k \ket{E_k}$ is prepared, where $E_k$ denotes the eigenstates and energies of the molecular Hamiltonian, the algorithm results in measurement of $E_k$ (as well as the preparation of $\ket{E_k}$) with probability $|c_k|^2$. So, the success of the procedure will depend on the overlap $c_k$ of the trial function with the targeted state. In Fig.~\ref{fig:results_variances} we show absolute values of the variances $\text{Var}(U)=|\expval{H^2}_U - \expval{H}_U^2|$ with $U\in\left\{U_\text{HF}, U_\text{SPA}\right\}$ as a quantifier for closeness to an eigenstate for BeH$_2$(4,8) and N$_2$(6,12). For both systems, SPA gives a significantly improved trial state and requires only depth 3 circuits with 6, respectively 9, controlled-not gates in total (see Tab.~\ref{tab:gate_counts}). 
Note that for stretched geometries of BeH$_2$, the variances of both initialization methods become almost identical as both methods are comparably close to electronic eigenstates. Hartree{\textendash}Fock initialization would however not result in a clear preference for the ground state, as the trial wavefunction has similar overlap with an excited state (shown in the central plot of Fig.~\ref{fig:results_variances}).
If the expected energy range of the ground state is known, this deficiency could for example be overcome with the \textit{Philter} algorithm~\cite{jensen2020quantum}. The associated costs are however significantly larger than simply switching to an improved trial state in the form of $U_\text{SPA}$. Other improved strategies on quantum phase estimation like in Ref.~\cite{berry2018improved} could be employed with a SPA trial state boosting the overall success probability.\\

In this work we use the basis-set-free approach of Ref.~\cite{kottmann2020reducing} in order to determine the spatial orbitals that form the qubit Hamiltonian in a system-adapted bottom-up approach. Our approach is therefore independent of static basis sets and a first assessment of its numerical performance compared to those basis sets would be interesting. In Ref.~\cite{kottmann2020reducing} first comparisons were already performed where the basis-set-free approach allowed significant improvements in numerical accuracy. In Fig.~\ref{fig:reactions} we provide three further examples in the form of small chemical reactions, that are significantly larger systems than in Ref.~\cite{kottmann2020reducing} and further confirm improved accuracy with a directly determined MRA-PNO basis.\\

In all calculations we observed fast convergence of the optimizations of the SPA circuits that usually took 4-10 BFGS iterations for canonical arrangement of primitive excitations. In all cases, a single BFGS iteration, required a single energy and gradient evaluation. It was furthermore possible to initialize all angles to zero (\textit{i.e.} using Hartree{\textendash}Fock as a starting point) without reaching local minima or plateaus~\cite{mcclean2018barren} in the process. This indicates that the optimization of SPA circuits can be achieved routinely and cheap with gradient based methods. UpCCGSD and 2-UpCCGSD behaved similar at points where it resulted in similar energies as with the SPA but became more difficult for stretched geometries. Here we needed to run several (5-10) optimizations with different starting points in order to achieve good convergence. With this strategy the optimization also took substantially more BFGS iterations (up to 100) which is however still a comparably small number~\cite{anselmetti2021local, jones2019variational, cervera2020meta}.\\

\section{Implementation}
All our circuit construction schemes are implemented in the \textsc{tequila}~\cite{tequila} library that also contains a convenient interface to \textsc{madness}~\cite{harrison2016madness} where the orbitals are computed according to Ref.~\cite{harrison2004multiresolution, bischoff2014regularizing} (Hartree{\textendash}Fock) and Ref.~\cite{kottmann2020direct} (MRA-PNO-MP2) with the standared (non-regularized) nuclear and electronic potentials. 
A specific point of the LiH(4,16) molecule of Fig.~\ref{fig:lih_and_c2h6} can for example be computed as 
\begin{lstlisting}[language=Python]
import tequila as tq

geometry="Li 0.0 0.0 0.0\n
          H 0.0 0.0 1.5"
mol = tq.Molecule(geometry,n_qubits=16)
H = mol.make_hamiltonian()
U = mol.make_ansatz(name="SPA")
U += mol.make_ansatz(name="GS", include_reference=False)
E = tq.ExpectationValue(H=H, U=U)

result = tq.minimize(E, method="bfgs")
\end{lstlisting}
where \texttt{make\_ansatz} automatically compiles circuits according to Eq.~\eqref{eq:pno_upccd} and adds the remaining unitary excitations.
It accepts and interprets all keywords  assembled from \texttt{name="\{HCB\}-\{SPA\}-UpCC\{G\}\{A\}\{S\}\{D\}"} where \texttt{HCB} will result in the circuit being compiled entirely in the HCB representation (meaning that $U_\text{HCB}^\text{JW}$ in Eq.~\eqref{eq:pno_upccd} will be removed), \texttt{SPA} will restrict all excitations to the surrogates excitation pattern (\textit{i.e.} excitations are restricted within the ${S}_k$ orbital sets), \texttt{D} and \texttt{S} will include doubles and singles, \texttt{A} will result in approximated singles as qubit excitations and \texttt{G} will result in generalized singles and doubles.
The additional keyword \texttt{direct\_compiling="ladder"} will result in the laddered arrangement of the SPA (see Figs.~\ref{fig:circuit_compiling_cartoon} and~\ref{fig:beh2_circuit}).
Invalid combinations, like the combination of \texttt{HCB} and \texttt{S} will result in exceptions. Note, that the \texttt{UpCC} part is not necessary in the name, but can be included to enhance readability in the code. The standard method is \texttt{SPA} which corresponds to Eq.~\eqref{eq:pno_upccd} and is equivalent to \texttt{SPA-UpCCD}. In this sense, \texttt{SPA-UpCCGD} would result in the SPA circuit complemented with all unaccounted generalized double excitations within the ${S}_k$ orbital sets.
The frozen-core approximation, i.e., no correlation of the $N_\text{c}$ ($5N_\text{c}$) lowest orbitals of molecules with $N_\text{c}$ second (third) row atoms, is enabled by default as well as active-spaces that include only pairs represented by more than one (the Hartree{\textendash}Fock) orbital. The SPA energies for $N_2$(6,12) in Fig.~\ref{fig:results_beh2_bh3_n2} can for example be computed as
\begin{lstlisting}[language=Python]
import tequila as tq
mol = tq.Molecule("n2.xyz",n_qubits=12)
H = mol.make_hardcore_boson_hamiltonian()
U = mol.make_ansatz("HCB-SPA-UpCCD")
E = tq.ExpectationValue(H=H, U=U)
result tq.minimize(E)
\end{lstlisting}
where the active space is automatically constructed. Here, we exploited the fact, that  PNO occupation numbers in the surrogate model are largest for the three orbitals that correspond to the triple bond, so that with \texttt{n\_pno=3} three PNOs from those pairs are selected automatically. More complicated active-spaces can be specified over the \texttt{active\_orbitals} keyword where information about all orbitals from the surrogate can be obtained over \texttt{print(mol)}. In the last code snipped we also illustrated how to optimize the separable pair ansatz directly in the hard-core Boson representation that allows simulations with $\Nq=\No$ qubits.
In this work, we used \textsc{qulacs}~\cite{qulacs} as quantum simulation backend, \textsc{scipy}~\cite{scipy} as optimization backend, and the Jordan-Wigner implementation of \textsc{openfermion}~\cite{OpenFermion}. Gradient compilation for the BFGS optimization is performed by the automatically differentiable framework of Ref.~\cite{kottmann2020feasible} where gradients for the controlled-$R_y$ operations of the optimized circuits follow the same principle. All of those options correspond to the defaults which do not need to be explicitly specified and we refer to Ref.~\cite{tequila} and~\cite{kottmann2020feasible} for more details on how to use \textsc{tequila} \textit{e.g.} for the manual construction of circuits that can be combined with the \texttt{U} objects constructed above.\\
Energies from classical quantum chemistry methods can be computed through \textsc{tequila} interfaces to \textsc{psi4}~\cite{psi4} and \textsc{pyscf}~\cite{pyscf1} for example via \texttt{mol.compute\_energy("ccsd")}.

\section{Conclusion \& Outlook}
We formulated a physically motivated recipe to construct low-depth and local quantum circuits that are able to approximate large parts of the electronic correlation in electronic structure problems. Integrated in a, now classically simulable, variational quantum eigensolver we observed fast and robust convergence for all test systems.
If applied to a closed shell reference state, the resulting circuits prepare wavefunctions equivalent to the PNO-UpCCD circuits introduced in Ref.~\cite{kottmann2020reducing} with significantly reduced depth and overal gate counts from several hundred to low one to two digit figures.
Due to their naturally separated form, the associated wavefunctions can be represented with linear memory requirement with respect to the system size which allows to optimize the parameters of the low-depth circuit in a classical pre-computation step. This bypasses challenges in variational quantum eigensolvers like finite-shot sensitive gradients and high measurement cost. 
Due to the physically inspired construction we furthermore expect this model to be well behaved with gradient based optimization.
Within this work we observed fast convergence in a few epochs of the BFGS optimizer throughout all numerical computations without getting trapped in local minima or plateaus.
All these properties qualify this model to be a potential minimum benchmark that quantum algorithms have to outperform in order to claim any advantage over classical methods.
In this regard, BeH$_2$(4,8) and N$_2$(6,12) could be well suited test systems. Furthermore the SPA framework offers a robust and efficient way to generate physically reasonable approximations to ground states that can be employed in the development of other methods. This was for example done in the context of explicitly correlated corrections~\cite{schleich2021improving} and bounds on true eigenvalues~\cite{weber2021reliability}.\\
Within quantum algorithms for electronic structure, we see the separable pair ansatz as initial part of larger approaches which we illustrated within two scenarios. The first employs the optimized SPA circuits as initial parts of a larger variational algorithm, here illustrated within the $k$-UpCCGSD hierarchy. The second uses the SPA as significantly improved initial state for phase estimation.\\
In this work, we integrated our methodologies into the basis-set-free framework of Ref.~\cite{kottmann2020reducing}, which is not a necessity to compile the low-depth circuits, but allows to compute basis-set-independent energies with high numerical accuracy.
For weakly correlated reactions, this provides a good balance between the one- and many-body aspects of the electronic wavefunctions which we illustrated on small organic reactions.
Our current implementation does not exploit the properties of the SPA wavefunction completely but rather takes advantage of high-performance simulators like \textsc{qulacs}~\cite{qulacs} within the \textsc{tequila}~\cite{tequila} framework. It is however well suited for systems treated in this work.
In the future, specialized high-performance implementations would be desirable and the combination with basis-set-free approaches could be interesting as a classical algorithm for weakly correlated molecular structures as they for example occur in a wide range of organic reactions. 
Within this context, to further enhance the overall performance of the model, we expect improvements on the surrogate model that determines the orbital basis. Additionally, one can include orbital optimization, which allows optimized linear-combination within said orbital basis. In the cases of C$_2$H$_6$ and LiH we demonstrated that with optimized orbitals, the SPA is able to describe the singe bond reliably through all bond distances. We expect this behaviour to remain consistent within other single bonds.
As the SPA model is classically simulable it sets a new benchmark for variational quantum algorithms. In the future it would be interesting to see how far this concept can be extended and if low-depth and local approaches that go beyond the classically tractable regime can be constructed on-top of an SPA initial state.

\section{Acknowledgement}
We would like to thank Philipp Schleich for providing valuable suggestions and comments on the manuscript and Garnet Kin-Lic Chan for pointing out similarities with generalized valence bond models.
This work was supported by the U.S. Department of Energy under Award No. DE-SC0019374.
A.A.-G. acknowledges the generous support from Google, Inc.  in the form of a Google Focused Award.
A.A.-G. also acknowledges support from the Canada Industrial Research Chairs  Program and the Canada 150 Research Chairs Program. Computations were performed on the niagara supercomputer at the SciNet HPC Consortium.~\cite{niagara1, niagara2} SciNet is funded by: the Canada Foundation for Innovation; the Government of Ontario; Ontario Research Fund - Research Excellence; and the University of Toronto.
We thank the generous support of Anders G. Fr\o{}seth.

\bibliography{main}
\onecolumngrid
\clearpage
\appendix
\section*{Appendix}

The following snipped of code reproduces results from Fig.~\ref{fig:spa_mra_vs_gbs} for BeH$_2$(4,8) with MRA-PNOs. The necessary dependencies are: \textsc{tequila} $\geq$v1.6 (github.com/tequilahub/tequila), \textsc{madness} (fork from 
github.com/kottmanj/madness, branch:\textit{tequila}, revision: \textit{827b714aa8cd737e33a7cac6de0ad403dcbb5b0b}), and \textsc{pyscf} $\geq$v1.7.6 (needed for convenient FCI computation only).
\begin{lstlisting}
import tequila as tq
import numpy, time

geometry="Be 0.0 0.0 0.0\nH 0.0 0.0 1.5\nH 0.0 0.0 -1.5"

start = time.time()
mol = tq.Molecule(geometry=geometry, n_qubits=8)
U = mol.make_ansatz("SPA")
print(U)
H = mol.make_hamiltonian()
E = tq.ExpectationValue(H=H, U=U)
result=tq.minimize(E)
stop = time.time()

print("{:25}:{:+2.5f}".format("SPA/MRA-PNO(4,8)", result.energy))
print("{:25}:{:+2.5f}".format("FCI/MRA-PNO(4,8)", mol.compute_energy("fci")))
\end{lstlisting}

In the same manner, the next snipped reproduces BeH$_2$(4,8) with Gaussian basis sets. Fig.~3 of Ref.~\cite{pyscf2} gives more details about the underlying \textsc{pyscf} submodule embodied in \textsc{tequila}s \texttt{optimize\_orbitals} function. Necessary dependencies are: \textsc{tequila} $\geq$v1.6.4 (github.com/tequilahub/tequila) and \textsc{pyscf} $\geq$v1.7.6. 
\begin{lstlisting}
import tequila as tq
import numpy, time

# pair 1
U = tq.gates.X(target=(0,))
U += tq.gates.Ry(target=(4,), angle="a")
U += tq.gates.X(target=(0,), control=(4,))

# pair 2
U += tq.gates.X(target=(2,))
U += tq.gates.Ry(target=(6,), angle="a")
U += tq.gates.X(target=(2,), control=(6,))

# HCB to JW
U += tq.gates.X(target=(1,), control=(0,))
U += tq.gates.X(target=(3,), control=(2,))
U += tq.gates.X(target=(5,), control=(4,))
U += tq.gates.X(target=(7,), control=(6,))

geometry="be 0.0 0.0 0.0\nh 0.0 0.0 1.5\nh 0.0 0.0 -1.5"

# better guess than starting from canonicals
x=1.0/numpy.sqrt(2)
mo_coeff=numpy.asarray([[x,x],[-x,x]])

for basis_set in ["STO-3G", "6-31G", "cc-pVDZ"]:

    start = time.time()

    mol = tq.Molecule(geometry=geometry, basis_set=basis_set)
    # freeze core orbital
    a,b,c=mol.get_integrals()
    mol = tq.Molecule(geometry=geometry, one_body_integrals=b, two_body_integrals=c, nuclear_repulsion=a, active_orbitals=list(range(1,mol.n_orbitals)))

    guess = numpy.eye(mol.n_orbitals)
    guess[0:mo_coeff.shape[0], 0:mo_coeff.shape[1]] = mo_coeff

    # optimize orbitals (pyscf module in the background)
    result1 = tq.chemistry.optimize_orbitals(molecule=mol, circuit=U, silent=True, initial_guess=guess)

    stop = time.time()

    # compute FCI(2,4) in the two optimized SPA orbitals
    a,b,c=result1.molecule.get_integrals()
    opt_mol = tq.Molecule(geometry=geometry, one_body_integrals=b, two_body_integrals=c, nuclear_repulsion=a, active_orbitals=[0,1,2,3])
    fci = numpy.linalg.eigvalsh(opt_mol.make_hamiltonian().to_matrix())[0]

    # use old result as starting guess for new basis_set (faster)
    mo_coeff = result1.mo_coeff

    print("{:25}:{:+2.5f}".format("SPA/{}(4,8)".format(basis_set), result1.energy))
    print("{:25}:{:+2.5f}".format("FCI/{}(4,8)".format(basis_set), fci))
    print("walltime: {}s".format(stop-start))
\end{lstlisting}
All wall times in Fig.~\ref{fig:spa_mra_vs_gbs} were measured on a Xeon(R) W-2135 CPU, 3.70GHz with 12 cores using \textsc{qulacs} as circuit simulation backend.

\end{document}